\documentclass[journal,transmag]{IEEEtran}
\usepackage{cite}
\usepackage{graphicx}
\usepackage{amssymb,amsmath}
\usepackage{color}
\usepackage{multicol}

\begin{document}

\title{Global biasing using a Hardware-based artificial Zeeman term in Spinwave Ising Machines}

\author{
	\IEEEauthorblockN{Victor H. Gonz\'alez\IEEEauthorrefmark{1}, Artem Litvinenko\IEEEauthorrefmark{1}, Roman Khymyn\IEEEauthorrefmark{1},  and Johan \AA kerman\IEEEauthorrefmark{1}}
	\IEEEauthorblockA{\IEEEauthorrefmark{1}Department of Physics, University of Gothenburg, Gothenburg, 41296, Sweden,\\  victor.gonzalez@physics.gu.se and johan.akerman@physics.gu.se}
}

\IEEEtitleabstractindextext{%
\begin{abstract}
A spinwave Ising machine (SWIM) is a newly proposed type of time-multiplexed hardware solver for combinatorial optimization that employs feedback coupling and phase sensitive amplification to map an Ising Hamiltonian into phase-binarized propagating spin-wave RF pulses in an Yttrium-Iron-Garnet (YIG) film. In this work, we increase the mathematical complexity of the SWIM by adding a global Zeeman term to a 4-spin MAX-CUT Hamiltonian using a continuous external electrical signal with the same frequency as the spin pulses and phase locked with with one of the two possible states. We are able to induce ferromagnetic ordering in both directions of the spin states despite antiferromagnetic pairwise coupling. Embedding a planar antiferromagnetic spin system in a magnetic field has been proven to increase the complexity of the graph associated to its Hamiltonian and thus this straightforward implementation helps explore higher degrees of complexity in this evolving solver.  
\end{abstract}

}

\maketitle

\pagestyle{empty}
\thispagestyle{empty}

\IEEEpeerreviewmaketitle


\IEEEPARstart{I}{n the landscape} of physical computation schemes, Ising machines (IM) have attracted considerable attention and investment over the last decade, in both academic and industrial research\cite{mohseni2022IMreview,litvinenko2022SWIM, albertsson2021ultrafastSHNOIM,houshang2022SHNOIM,Honjo2021SciAdv100kCIM,tatsumura2021SBFPGA}, for their applicability to combinatorial optimization problems, potential for scalability and progressive increase in mathematical complexity. In this work, we contribute to the latter as we implement a global bias to artificial spin states in a spinwave Ising machine (SWIM). 

A SWIM is a newly proposed~\cite{litvinenko2022SWIM} time-multiplexed hardware solver circuit for problems in the NP (nonpolynomial time) complexity class that employs feedback coupling and phase sensitive amplification to map an Ising Hamiltonian into spinwave (SW) RF pulses propagating in an Yttrium-Iron-Garnet (YIG) film. Spinwaves are suitable for Ising machines because of their GHz oscillation frequencies, which permits the development of multiphysical systems using cheap and efficient off-the-shelf microwave components for signal processing \cite{litvinenko2021tunable, litvinenko2018chaotic} and amplification which results in a small circuitry footprint and high per-spin power efficiency. In this work, we use an external continuous microwave signal to implement a global biasing to propagating spinwave RF artificial spin states. Exploring the complexity limits of this circuit implementation thus holds significant interest for technical and commercial applications.

An IM operates as an mapping of an objective function into the Ising Hamiltonian of a device composed of an array of $N$ binarized physical units referred to as spins $s_i=\pm 1$: 
\begin{equation}
    H = -\frac{1}{2} \sum_{i=1}^N \sum_{j=1}^N J_{ij} s_i s_j - \sum_{i=1}^N h_is_i
    \label{eq:hamiltonian}
\end{equation}

The objective function associated with the NP problem to solve is encoded into the pairwise coupling $J_{ij}$ and external Zeeman bias $h_i$ such that ground state of the system represents the solution to it. Combinatorial problems of practical use, such as the traveling salesman and knapsack with integer weights, have been shown to be encodable if one can add constraints to the degrees of freedom of the system \cite{lucas2014ising}. Adding a global bias (i.e. the $h_i$ is identical for all $s_i$) to an antiferromagnetic planar graph has been shown as a straightforward way to increase its mathematical complexity~\cite{Barahona1982}. Implementation of a global Zeeman term using software has also shown to improve the stability and performance of time-multiplexed IMs with frustated lattices \cite{Inui2022}. Thus, exploring hardware-based alternative schemes can lend versatility to small scale low-power devices.

\begin{figure} 
    \centering
    \includegraphics[width=\linewidth]{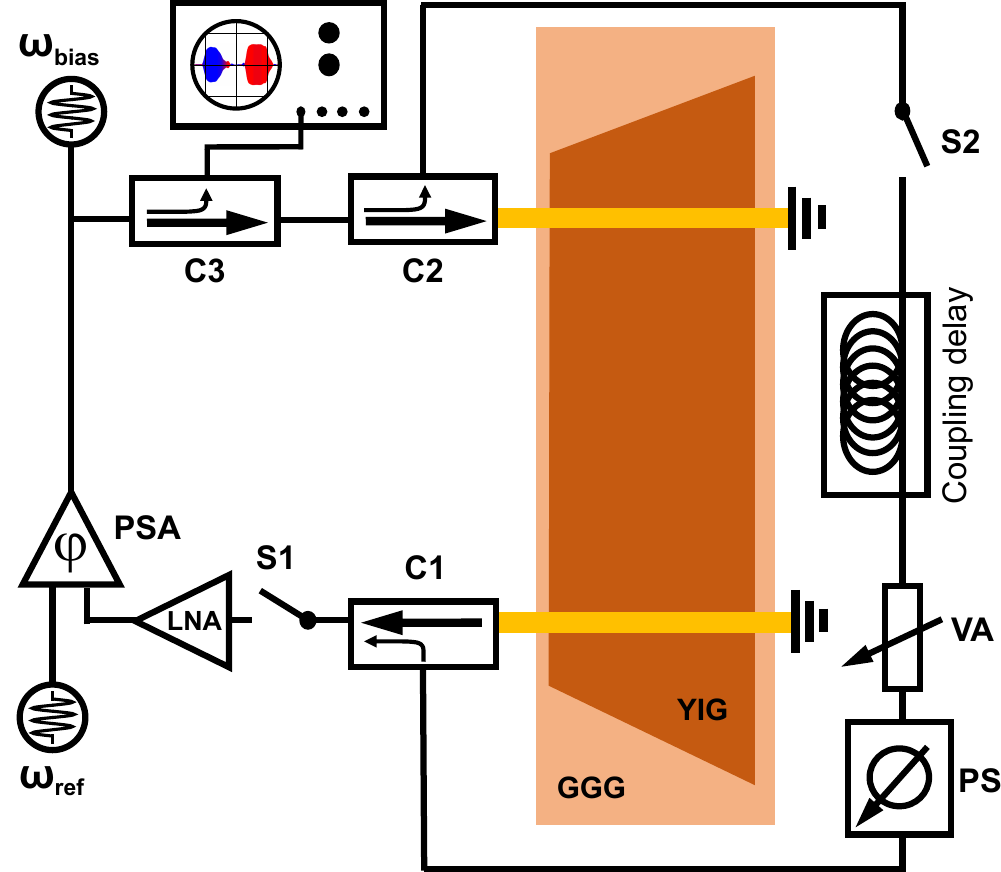}
    \caption{Zeeman-biased spinwave Ising machine. PSA and LNA stand for phase-sensitive and low noise amplifiers, respectively. The propagating RF pulses have frequency of 3.13 GHz. The sign of the coupling is controlled by the total phase accumulation in the coupling delay and the Zeeman field amplitude and sign is controlled by the amplitude and phase of the injected signal $\omega_{bias}$.}
    \label{fig:circuit-layout}
\end{figure}

\begin{figure*}[ht!]
    \centering
    \includegraphics[width=\textwidth]{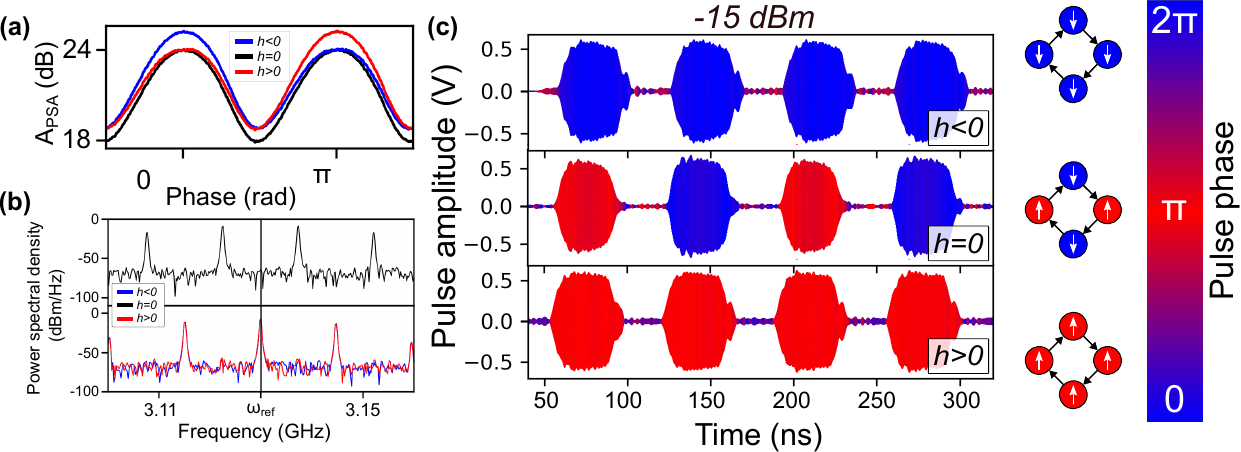}
    \caption{Influence of Zeeman term at different signs of magnetic field. (a) Measured potential landscape shift of the artificial spin states. The sign of $h$ is given by the phase between $\omega_{ref}$ and $\omega_{bias}$. The total phase sensitive amplification of the circuit $A_{PSA}$ favors either 0 or $\pi$ depending on $h$ and thus changes the magnetic ordering of the spins. (b) Power spectral densities (PSD) as a functions of frequency for different $h$ signs. We observe that the modulation harmonics can be used to identify the type of solution achieved, with both ferromagnetic states having the same spectra. (c) Time traces of the RF pulses colored with their respective instantaneous phase for different signs of $h$. The non-zero signals have an amplitude of -15~dBm. $\omega_{bias}$ allows us to change the ordering and direction of the artificial spin state.} 
    \label{fig:chage-of-ordering}
\end{figure*}

The circuit of the SWIM with a hardware implemented Zeeman term is shown in Fig.~\ref{fig:circuit-layout}. The SWIM's construction of the artificial spin state relies in phase-sensitive amplification (PSA) of a reference signal $\omega_{ref}=$3.13~GHz. The PSA binarizes the phase of the signal by amplifying only its in-phase ($\phi=0$) and out-of-phase ($\phi=\pi$) components. The signal is then simultaneously injected into a YIG waveguide and a coaxial delay line using coupler C2. The electrical signal excites spin waves within waveguide that propagate at a much slower speed, allowing the coupling delay to re-inject a shifted signal using coupler C1. Switch 1 (S1) then pulses the signal and a low noise amplifier (LNA) compensates propagation losses as the cycle starts again. The resulting time-multiplexed pulse train is our artificial spin state, where each pulse is an artificial spin, with their electrical amplitudes representing the norm of $s_i$ and their individual phases representing its sign. The coupling delay's length is such that every pulse interferes (or couples) with the previous nearest neighbor. Phase shifter PS ensures that the coupling term $J_{i,i+1}$ is antiferromagnetic and variable attenuator VA controls its strength. The Zeeman term is implemented with external signal $\omega_{bias}$ applied to the propagating pulses after PSA.

The role of $\omega_{bias}$ is to unbalance the potential landscape of the artificial spin states and favor one phase over the other. Fig.\ref{fig:chage-of-ordering}(a) shows the changes in PSA amplitude ($A_{PSA}$) for different signs of h. The effective sign of the Zeeman term is given by its relative phase with respect to the reference signal, with negative $h$ being in-phase and positive out-of-phase. The effective magnitude of $h$ is given by the signals' amplitude, -15 dBm in this case. The amplification imbalance allows us to change the phase sensitivity of the circuit and globally bias the state of the spins. The consequences of the bias are shown in Fig.\ref{fig:chage-of-ordering}(b) and (c). In (b), we observe that the modulation harmonics present in the spectra depend on the spin state, with the biased signal containing a central carrier corresponding to $\omega_{ref}$ that is absent for the unbiased solutions. $\omega_{bias}$, whose frequency is the same as the reference signal's, drives the oscillators as they all acquire the same phase (i.e. align their spin direction).  We complement this picture in (c) using the time traces of the RF pulses colored with their respective instantaneous phase and associated graph. It is clear that despite antiferromagnetic coupling, we are able to induce a change in ordering in both possible directions using $\omega_{bias}$ alone. Combining time trace and spectrum analyses, we have different tools to study the effectiveness of the biasing as well as develop differentiation and operation protocols that allow us to program more complex problems. Although this is very promising in terms of exploration of higher complexity schemes,  unexpected states also appear for intermediate amplitudes of $h$.

\begin{figure}
    \centering
    \includegraphics[width=\linewidth]{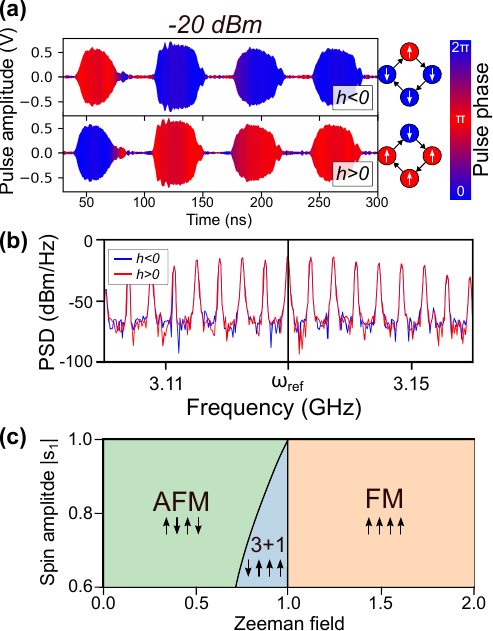}
    \caption{Mixed artificial spin states for $\omega_{bias}$ with -20~dBm. (a) Colored time traces of the RF spins. The appearance of 3+1 states is evidence of additional degrees of freedom in the Hamiltonian of the system. (b) Phase diagram of a ring-shaped 4 spin Ising machine with variable spin amplitude. The 3+1 states are a consequence of phase-dependent amplification of the spins. (c) PSD for both solutions with 3+1 states, we observe the same synchronization as in fig.\ref{fig:chage-of-ordering}(b), but the modulation peaks are characteristic to these solutions.}
    \label{fig:3+1-states}
\end{figure}

While it is expected that there will be a sudden breaking of up and down spin symmetry at $h=2$, a gradual spin flipping occurs. Mixed spin states appear at intermediate values of $\omega_{bias}$ amplitude ($\approx$-20 dBm), as seen in Fig.\ref{fig:3+1-states}(a), resembling a chain with magnetic domains. Although the system is indeed synchronized with $\omega_{bias}$, as shown in fig.\ref{fig:3+1-states}(b), these 3+1 states (three spins up and one down, or vice versa) are not a minimum energy solution of eq.\ref{eq:hamiltonian} and suggest an unintended increase of the degrees of freedom of the system.

In Fig.\ref{fig:3+1-states}(c) we show the phase diagram for a ring-shaped 4 spin system under an external Zeeman field where one of the spin's amplitude $|S_1|$ can be less than one, i.e. one spin is shorter. The shortening of $S_1$ results in a phase transition of the 4-spin system and appearance of the 3+1 states. In fact, from the time traces shown in Fig.\ref{fig:3+1-states}, we can directly observe that the spins have different electrical amplitudes and the odd one out has the smallest one. Since electrical amplitude and spin amplitude are proportional to each other, the emergence of the 3+1 states can give us information about the operation of our circuit and its limitations. 

Probing into the origin of this amplitude mismatch, we propose that the emergence of the 3+1 states depends on the non-linearity of the saturation of the LNA. Since we are injecting additional power to the system with the Zeeman signal, the LNA saturates and thus gain compression occurs beyond its linear range. Even if they are very close in the power transfer curve\cite{LNAdatasheet}, its non-linearity results in spins of different signs being amplified differently with the minority spin shortening. An amplifier with a higher linear regime would mitigate this state degeneration. If compression gain is unavoidable, a stronger coupling between spins and a smoother saturation curve for the LNA would suppress 3+1 states. Digital feedback using a field programmable gate array (FPGA) has been employed successfully in previous time-multiplexed Ising machines to improve amplitude stability \cite{Takesue2020} and can also be used instead to the delay line to modify pairwise or all-to-all coupling. These findings can be implemented in future circuit designs to improve the quality and complexity of the solutions with larger amounts of spins.

\begin{figure}
    \centering
    \includegraphics[width=\linewidth]{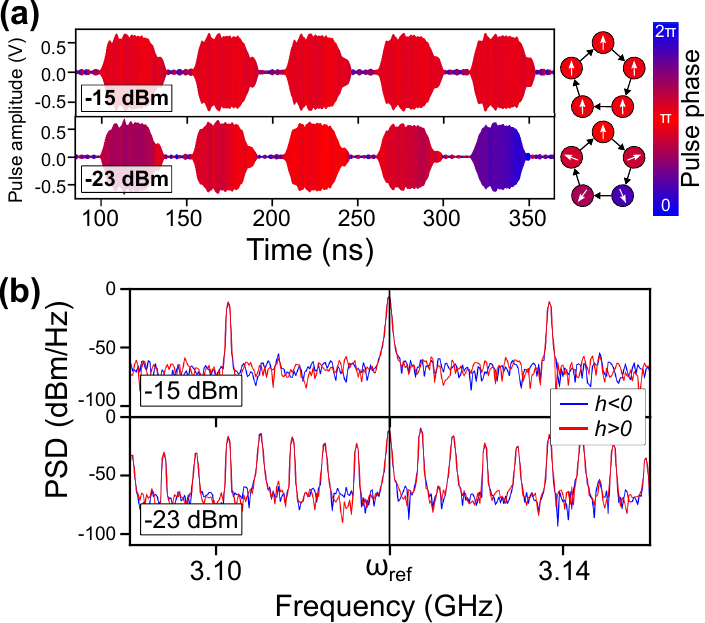}
    \caption{Five spin states at two different $\omega_{bias}$ amplitudes. (a) Time traces at -15 dBm and -20 dBm. We can observe that although the bias manages to stabilize the phase, the solution is not phase binarized. (b) Power spectral densities (PDS) of the solutions at the same amplitudes. We see that each solution is synchronized to $\omega_{bias}$ has a characteristic spectrum.}
    \label{fig:5-spins}
\end{figure}

It is worthwhile to mention that spectral analysis can help us understand the synchronization dynamics of the system as well as use for soution differentiation. As we see in fig.\ref{fig:chage-of-ordering}(b) and fig.\ref{fig:3+1-states}(b), the peaks at $\omega_{ref}$ show that $\omega_{bias}$ drives the oscillators. Additionally, both biased solutions have their own characteristic spectrum. Thus, including this information in the digital feedback can allow us to design differentiation metrics and stopping conditions for relaxation and annealing protocols in bigger and more complex systems. 

Finally, we tried to recreate a stable globally biased 3+2 solution with five spins as it would appear in a spin ring. Evidently, a ring with an odd number of spins will not have a stable unbiased solution as the phase difference on each circulation period will never be zero. It was clear that for a large enough bias, we do see all spins parallel to each other, as shown for -15 dBm in fig.\ref{fig:5-spins}(a). The phase transition mentioned before and shown in fig.\ref{fig:3+1-states} could be an indication that a 3+2 state could be viable and could allow us to construct a magnetic system analog with two clearly defined domains. An amplitude of -23 dBm  (fig.\ref{fig:5-spins}(a)), is unable to produce such solution because the spins do not synchronize with the external field (we can observe that the highest peak in fig.\ref{fig:5-spins}(b) is not at $\omega_{ref}$). Instead, the resulting state's phase is not binarized and produces spins do not comply with the definition of eq.\ref{eq:hamiltonian}. We believe that frustration is responsible for this phase slip and stable solutions are achievable with digital coupling and individual bias, both implementable with the aforementioned FPGA.

We have shown the broad features of a globally biased artificial spin state space composed of RF pulses in a YIG waveguide. Despite  antiferromagnetic coupling between nearest neighbors in a 4 spin ring, we are able to induce ferromagnetic ordering by injecting an external  signal of the same frequency to emulate the role of the Zeeman term in the Ising Hamiltonian. Intermediate values of this Zeeman signal introduce degeneracy in the amplification of the pulses and, consequently, 3+1 spin states. These effects be mitigated by alternative amplification schemes whose implementation would guide future work in enabling all-to-all spin coupling for tackling non-trivial optimization tasks. The present work improves upon the emerging technology of commercially feasible IM hardware accelerators. The SWIM concept has a high potential for further scaling in terms of spin capacity, physical size and low-power low-footprint circuits for applied combinatorial optimization.
\bibliographystyle{IEEEtran}
\bibliography{IEEEabrv,refs}

\begin{thebibliography}{10}
\providecommand{\url}[1]{#1}
\csname url@samestyle\endcsname
\providecommand{\newblock}{\relax}
\providecommand{\bibinfo}[2]{#2}
\providecommand{\BIBentrySTDinterwordspacing}{\spaceskip=0pt\relax}
\providecommand{\BIBentryALTinterwordstretchfactor}{4}
\providecommand{\BIBentryALTinterwordspacing}{\spaceskip=\fontdimen2\font plus
\BIBentryALTinterwordstretchfactor\fontdimen3\font minus
  \fontdimen4\font\relax}
\providecommand{\BIBforeignlanguage}[2]{{%
\expandafter\ifx\csname l@#1\endcsname\relax
\typeout{** WARNING: IEEEtran.bst: No hyphenation pattern has been}%
\typeout{** loaded for the language `#1'. Using the pattern for}%
\typeout{** the default language instead.}%
\else
\language=\csname l@#1\endcsname
\fi
#2}}
\providecommand{\BIBdecl}{\relax}
\BIBdecl

\bibitem{mohseni2022IMreview}
N.~Mohseni, P.~L. McMahon, and T.~Byrnes, ``Ising machines as hardware solvers
  of combinatorial optimization problems,'' \emph{Nature Reviews Physics},
  vol.~4, no.~6, pp. 363--379, 2022.

\bibitem{litvinenko2022SWIM}
A.~Litvinenko, R.~Khymyn, V.~H. Gonz{\'a}lez, A.~A. Awad, V.~Tyberkevych,
  A.~Slavin, and J.~{\AA}kerman, ``A spinwave ising machine,'' \emph{arXiv
  preprint arXiv:2209.04291}, 2022.

\bibitem{albertsson2021ultrafastSHNOIM}
D.~I. Albertsson, M.~Zahedinejad, A.~Houshang, R.~Khymyn, J.~{\AA}kerman, and
  A.~Rusu, ``Ultrafast ising machines using spin torque nano-oscillators,''
  \emph{Applied Physics Letters}, vol. 118, no.~11, p. 112404, 2021.

\bibitem{houshang2022SHNOIM}
A.~Houshang, M.~Zahedinejad, S.~Muralidhar, R.~Khymyn, M.~Rajabali, H.~Fulara,
  A.~A. Awad, J.~{\AA}kerman, J.~Checi{\'n}ski, and M.~Dvornik,
  ``Phase-binarized spin hall nano-oscillator arrays: Towards spin hall ising
  machines,'' \emph{Physical Review Applied}, vol.~17, no.~1, p. 014003, 2022.

\bibitem{Honjo2021SciAdv100kCIM}
T.~Honjo, T.~Sonobe, K.~Inaba, T.~Inagaki, T.~Ikuta, Y.~Yamada, T.~Kazama,
  K.~Enbutsu, T.~Umeki, R.~Kasahara, K.~ichi Kawarabayashi, and H.~Takesue,
  ``100,000-spin coherent ising machine,'' \emph{Science Advances}, vol.~7,
  no.~40, p. eabh0952, 2021.

\bibitem{tatsumura2021SBFPGA}
K.~Tatsumura, M.~Yamasaki, and H.~Goto, ``Scaling out ising machines using a
  multi-chip architecture for simulated bifurcation,'' \emph{Nature
  Electronics}, vol.~4, no.~3, pp. 208--217, 2021.

\bibitem{litvinenko2021tunable}
A.~Litvinenko, R.~Khymyn, V.~Tyberkevych, V.~Tikhonov, A.~Slavin, and
  S.~Nikitov, ``Tunable magnetoacoustic oscillator with low phase noise,''
  \emph{Physical Review Applied}, vol.~15, no.~3, p. 034057, 2021.

\bibitem{litvinenko2018chaotic}
A.~Litvinenko, S.~Grishin, Y.~P. Sharaevskii, V.~Tikhonov, and S.~Nikitov, ``A
  chaotic magnetoacoustic oscillator with delay and bistability,''
  \emph{Technical Physics Letters}, vol.~44, no.~3, pp. 263--266, 2018.

\bibitem{lucas2014ising}
A.~Lucas, ``Ising formulations of many np problems,'' \emph{Frontiers in
  physics}, vol.~2, p.~5, 2014.

\bibitem{Barahona1982}
\BIBentryALTinterwordspacing
F.~Barahona, ``On the computational complexity of ising spin glass models,''
  \emph{Journal of Physics A: Mathematical and General}, vol.~15, no.~10, p.
  3241, oct 1982. [Online]. Available:
  \url{https://dx.doi.org/10.1088/0305-4470/15/10/028}
\BIBentrySTDinterwordspacing

\bibitem{Inui2022}
\BIBentryALTinterwordspacing
Y.~Inui, M.~D. S.~H. Gunathilaka, S.~Kako, T.~Aonishi, and Y.~Yamamoto,
  ``Control of amplitude homogeneity in coherent ising machines with artificial
  zeeman terms,'' \emph{Communications Physics}, vol.~5, no.~1, p. 154, Jun
  2022. [Online]. Available: \url{https://doi.org/10.1038/s42005-022-00927-x}
\BIBentrySTDinterwordspacing

\bibitem{LNAdatasheet}
\BIBentryALTinterwordspacing
MiniCircuits. Coaxial low noise amplifier zx60-83ln12+. [Online]. Available:
  \url{https://www.minicircuits.com/pdfs/ZX60-83LN12+.pdf}
\BIBentrySTDinterwordspacing

\bibitem{Takesue2020}
\BIBentryALTinterwordspacing
H.~Takesue, K.~Inaba, T.~Inagaki, T.~Ikuta, Y.~Yamada, T.~Honjo, T.~Kazama,
  K.~Enbutsu, T.~Umeki, and R.~Kasahara, ``Simulating ising spins in external
  magnetic fields with a network of degenerate optical parametric
  oscillators,'' \emph{Phys. Rev. Appl.}, vol.~13, p. 054059, May 2020.
  [Online]. Available:
  \url{https://link.aps.org/doi/10.1103/PhysRevApplied.13.054059}
\BIBentrySTDinterwordspacing

\end{thebibliography}

\end{document}